\documentclass[usenatbib,useAMS]{mn2e}

\usepackage{amssymb,amsmath}
\usepackage{graphicx}
\usepackage{epsfig}
\usepackage{bbm}

\usepackage{color}

\newcommand{\beq}{\begin{equation}}
\newcommand{\eeq}{\end{equation}}
\newcommand{\bea}{\begin{eqnarray}}
\newcommand{\eea}{\end{eqnarray}}
\newcommand{\vc}[1]{{\textbf{#1}}}

\newcommand{\mc}[1]{\mathcal{#1}}

\newcommand{\dd}{{\rm d}}
\newcommand{\fett}[1]{\boldsymbol{#1}}

\newcommand{\nabq}{\boldsymbol{\nabla}_{\boldsymbol{q}}}

\begin{document}

\title[On the accuracy of N-body simulations at very large scales]{On the accuracy of N-body simulations at very large scales}

\author[G. Rigopoulos and W. Valkenburg]{Gerasimos Rigopoulos$^a$\thanks{email address: rigopoulos@thphys.uni-heidelberg.de} and Wessel Valkenburg$^b$\thanks{email address: valkenburg@lorentz.leidenuniv.nl}\\
$^a$ Institut f\"ur Theoretische Physik, Philosophenweg 12,\\
Universit\"at Heidelberg, 69120 Heidelberg, Germany\\
$^b$Instituut-Lorentz for Theoretical Physics, Universiteit Leiden, Niels Bohrweg 2, Leiden, NL-2333 CA, The Netherlands}

\maketitle

\begin{abstract}
{We examine the deviation of Cold Dark Matter particle trajectories from the Newtonian result as the size of the region under study becomes comparable to or exceeds the particle horizon. To first order in the gravitational potential, the general relativistic result coincides with the Zel'dovich approximation and hence the Newtonian prediction on all scales. At second order, General Relativity predicts corrections which overtake the corresponding second order Newtonian terms above a certain scale of the order of the Hubble radius. However, since second order corrections are very much suppressed on such scales, we conclude that simulations which exceed the particle horizon but use Newtonian equations to evolve the particles, reproduce the correct trajectories very well. The dominant relativistic corrections to the power spectrum on scales close to the horizon are at most of the order of $\sim  10^{-5}$ at $z=49$ and $\sim 10^{-3}$ at $z=0$. The differences in the positions of real space features are affected at a level below $10^{-6}$ at both redshifts. Our analysis also clarifies the relation of N-body results to relativistic considerations.
}
\end{abstract}

\section{Introduction}

As ongoing and future Large Scale Structure surveys will be mapping significant fractions of the observable Universe around our position, there is a corresponding trend for performing N-body simulations of increasing size - see \citep{alimi:2012be,kim:2011ab,2013arXiv1310.3740K} for the largest current simulations. Assessing the statistical significance of extremely massive haloes or accurate modeling of the fluctuation power at large scales, crucial for Baryonic Acoustic Oscillations and Dark Energy studies, call for large scale simulations in order to bring statistical uncertainties down to cosmic variance levels. As simulations are solving Newtonian dynamical equations, utilizing Newtonian gravity, one might be concerned about the validity of ultra large simulations since an increasing  box size eventually encompasses and exceeds the Hubble radius, particularly at the early times when initial conditions for the simulations are set up. For example, the initial displacements and velocities of particles are determined using the Zel'dovich approximation (ZA), or even second order Lagrangian Perturbation Theory (2LPT), which is a Newtonian solution. Further evolution is determined by solving entirely Newtonian equations. For large enough boxes these initial stages of evolution find the particles distributed on super-hubble scales, where relativistic effects would a priori be relevant and issues of the interpretation of coordinates (gauge issues) arise - see \citep{Chisari:2011iq, Green:2011wc, Flender:2012nq, Haugg:2012ng} for recent considerations. The box size of the largest simulations currently available is commensurable with the Hubble radius even at z=0.

In this paper we quantify the importance of relativistic corrections for N-body simulations at very large scales.\footnote{See \citep{Adamek:2013wja} for an N-Body approach that includes the leading order General Relativistic corrections on subhorizon scales.} Instead of using quantities such as the density or the metric functions and their power-spectra, we focus on the \emph{trajectories} of Cold Dark Matter (CDM) particles with respect to an appropriate coordinate frame, as this is a natural output of N-body simulations.  At leading order in the initial gravitational potential, the General Relativistic (GR) result for the trajectories coincides with the Zel'dovich approximation \citep{Russ:1995eu, Rampf:2012pu} and hence the Newtonian solution at this order \citep{Chisari:2011iq, Green:2011wc}.\footnote{See also \citep{Hwang:2012bi} for the Newtonian-Relativistic correspondence for cosmological perturbations in various gauges.} Relativistic corrections only appear at second order in the gravitational potential\footnote{In this work we do not include vectors and tensors. See \citep{Bruni:2013mua, Rampf:2013dxa} for a discussion of vector modes.} and overtake the second order Newtonian terms at large scales. However, all second order terms, including the relativistic ones, are much suppressed compared to the leading order Zel'dovich displacements at these scales. Hence, to this accuracy large N-body simulations reproduce the correct trajectories with respect to a Newtonian coordinate system.

In the next section we describe the framework in which we obtain the relativistic corrections to the CDM particles' trajectories. We start from a relativistic gradient expansion in a comoving synchronous frame which corresponds to a Lagrangian description of the dynamics on long wavelengths. The particle trajectories, including the relativistic contributions, are obtained via a transformation to a Newtonian frame. In section 3 we analyze the various contributions to the trajectories and quantify the scales at which the relativistic terms become dominant over the corresponding Newtonian terms. We close with a discussion in section 4 where we also touch upon the issue of gauge choice in relation to the coordinates employed in large simulations.

\section{Gradient expansion and particle trajectories}
The effect of long wavelength gravity fields on CDM particles can be described via a gradient expansion solution to the Einstein Equations \citep{Parry:1993mw,Comer:1994np, Rigopoulos:2012xj}. This approach starts by considering each different region of the universe evolving independently of its neighbouring regions. It then proceeds to take into account interactions of these different patches by including terms containing an increasing number of spatial derivatives. In particular, we start by writing the metric in the synchronous comoving frame
\begin{align}
 \label{co-synch}
 &\dd s^2 =-\dd t^2+\gamma_{ij}(t,\vc{q}) \, \dd q^{i} \dd q^{j}\,,
\end{align}
and solve for $\gamma_{ij}$ in a series of terms with an increasing number of spatial gradients. One then obtains the gradient expansion metric (GEM) \citep{Rigopoulos:2012xj}
\begin{align}
\gamma_{ij} &\simeq a^2\delta_{ij}\left[ 1+\frac{10}{3}\Phi(\vc{q})\right] \nonumber \\
  + & \frac{20}{3} \lambda(a)\left[\Phi_{,ij}\left(1-\frac{10}{3}\Phi\right)
 -5\Phi_{,i}\Phi_{,j}+\frac{5}{6}\delta_{ij}\Phi_{,l}\Phi_{,l} \right]\nonumber\\
+&\,T_1(a) \,\Phi_{,li}\Phi_{,lj} -T_2(a)\, \Phi_{,ll}\Phi_{,ij}\nonumber\\
-& \frac{T_2(a)}{4} \left[\Phi_{,lm}\Phi_{,lm}-\Phi_{,ll}\Phi_{,mm}\right]  \delta_{ij} \label{metric-phi}\,,
\end{align}
where the time independent $\Phi(\vc{q})$ parameterizes the initial metric perturbation and the time dependent functions satisfy
\bea
\frac{\dd J}{\dd a}+ \frac{J}{a}=\frac{1}{2aH}\,,\,\,\,\,
\frac{\dd \lambda}{\dd a}-2\frac{\lambda}{a} =\frac{J}{aH}\,,\,\,\,\,
\frac{\dd L}{\dd a}- \frac{L}{a}=\frac{J^2}{aH} \,.\!\!
\label{coeffs}\eea
with
\begin{align}
 T_1 &=a^2\int\limits^a_0 \frac{\dd x}{x^5H(x)}\frac{200}{3} \left[\lambda(x) J(x)-\frac{1}{3}L(x)\right] \,, \\
 T_2 &=a^2\int\limits^a_0 \frac{\dd x}{x^5H(x)}\frac{400}{9}\left[\lambda(x) J(x)-\frac{1}{2}L(x)\right] \,.
\end{align}
We have used the background FLRW scale factor $a(t)$ as the time variable and $ H(a) =H_0\sqrt{\Omega_M a^{-3}+\Omega_\Lambda}$. Note that the solution (\ref{metric-phi}) coincides with that found in \citep{Russ:1995eu} - see also \citep{Matarrese:1995sb, Matarrese:1997ay}.

The metric (\ref{metric-phi}) is determined up to initial conditions for the functions $J$, $\lambda$ and $L$. Setting them to zero at the initial time is equivalent to keeping only the fastest growing modes. At early enough times $a \rightarrow 0$, when the contribution from $\Lambda$ is negligible, they become
$J\simeq\frac{a^{3/2}}{5H_0\sqrt{\Omega_M}}$, $\lambda\simeq\frac{a^3}{5H_0^2\Omega_M}$, $L\simeq\frac{2}{175}\frac{a^{9/2}}{H_0^3\Omega_M^{3/2}}$
and $a\simeq t^{2/3}\left(H_0\sqrt{\Omega_M}\right)^{2/3}$. This also sets the initial extrinsic curvature to be homogenous $K^i_{j}\equiv\gamma^{il}\dot{\gamma}_{lj} \rightarrow 2H(t_{\rm in})\delta^i_j$ where $H(t_{\rm in})$ is the initial homogeneous Hubble rate. The CDM density is given by
\beq\label{density}
\rho(t,\vc{q})=\frac{3H_0^2\Omega_{M}}{8\pi G}\frac{\left(1+\frac{10}{3}\Phi(\vc{q})\right)^{3/2}}{\sqrt{{\rm Det}\gamma_{ij}(t,\vc{q})}}\left(1+\frac{\delta \rho_{\rm i}}{\bar{\rho}_{\rm i}}\right)\,,
\eeq
where $\frac{\delta \rho_{\rm i}}{\bar{\rho}_{\rm i}}$ is the initial fractional density perturbation determined by the energy constraint, see the treatment in \citep{Comer:1994np} and also \citep{Bruni:2013qta}. Ignoring the initial perturbation in (\ref{density}) recovers the fastest growing modes. Expanding ${\rm Det}\gamma_{ij}$ in (\ref{density}) to first order in $\Phi$, and ignoring terms subdominant at late times, returns
\beq\label{density-pert}
\frac{\delta\rho}{\bar{\rho}}\left(t,\vc{q}\right)\simeq -\frac{\lambda}{a^2}\frac{10}{3}\nabla^2_{\vc{q}}\Phi(\vc{q})
\eeq
which is the linear density perturbation in the synchronous gauge growing mode. It can be checked that $\frac{\lambda}{a^2}$ coincides with the density contrast growth factor once decaying modes become negligible. In order to link to the adiabatic inflationary initial conditions one simply notes that
\beq
1+\frac{10}{3}\Phi(\vc{q})={\exp}(2\zeta)
\eeq
where
\beq
\zeta=\zeta_{\rm G} + \frac{3}{5}f_{\rm NL}\zeta_{\rm G}^2\,,
\eeq
with $\zeta_{\rm G}$ the leading order gaussian perturbation from inflation.

The metric (\ref{metric-phi}) refers to a comoving coordinate system akin to a Lagrangian description of the dynamics. However, N-body simulations work in an Eulerian description where fluid elements (particles) move w.r.t a fixed coordinate system. It is therefore natural to ask how the above synchronous comoving description can be translated in terms directly comparable to the outcome of an N-body simulation: a set of Eulerian particle \emph{trajectories} under the influence of gravity. As a first step it is important to define what is meant by the coordinates used in an N-body simulation in relativistic terms.

The correspondence of the coordinates employed in an N-body simulation - the points of a Euclidean grid $\vc{x}$ and a universally ticking clock $\tau$ - to events in spacetime constitutes a choice of gauge.\footnote{A choice of gauge in General Relativity is a correspondence between some fictitious reference spacetime parameterized by a set of convenient coordinates and the true spacetime \citep{Bardeen:1980kt}. Here the fictitious spacetime is parameterized by the coordinates of the simulation which provide a realization of FRW spacetime.} The most natural choice from a simulator's perspective is the Newtonian gauge with the perturbed metric assigned to the simulation taking the Newtonian form
\begin{align}\label{Newton-metric}
g_{00}(\tau,\vc{x})&=-\left[ 1+2A(\tau,\vc{x}) \right] \,, \\
g_{0i}(\tau,\vc{x})&=0 \,, \\
g_{ij}(\tau,\vc{x})&=\delta_{ij}\left[ 1-2B(\tau,\vc{x}) \right ] a^2(\tau) \,,
\end{align}
where $A\ll 1$ and $B\ll 1$. The transformation between the synchronous and the Newtonian frame is determined by
\begin{align}
x^i(t,\vc{q})  &= q^{i}+\mc{F}^i(t,\vc{q})\label{xfn1} \,, \\
\tau (t,\vc{q})&= t+\mc{L}(t,\vc{q})\,.\label{xfn2}
\end{align}
Since a fixed value of $\vc{q}$ labels the worldline of a particle in the comoving frame, it follows that when $\vc{x}$ in (\ref{xfn1}) is expressed in terms of the Newtonian time $\tau$, instead of the particle's proper time $t$, it describes the trajectory in the Newtonian N-body frame of a particle with initial coordinate $\vc{q}$. At second order the trajectory reads \citep{Rampf:2013ewa, Rampf:2012pu}
\begin{align}
\label{x2}
\vc{x}(\tau,\fett{q}) &\simeq   \vc{q}  + \nabq S(\tau,\vc{q})
\end{align}
where
\begin{align}\label{x3}
&S=\frac{10}{3}\frac{\lambda(\tau)}{a^2(\tau)}\Phi(\vc{q}) +\frac{1}{8}\frac{T_2(\tau)}{a^2(\tau)}\frac{1}{\nabq^2}F \nonumber\\
&+\frac{50}{9}\frac{\left[ \lambda(\tau)+J^2(\tau)\right]}{a^2(\tau)}\frac{1}{\nabq^2}\left(\Phi_{,l}\Phi_{,l}-\frac{3}{2}\frac{1}{\nabq^2}F\right)\nonumber\\
&-\frac{50}{9}\frac{\left[ 2\lambda(\tau)+J^2(\tau) \right]}{a^2(\tau)}\Phi^2 \,,
\end{align}
where $1/\nabq^2$ denotes the inverse Laplacian and
\begin{align}
 F&= \Phi_{,lm}\Phi_{,lm}-\Phi_{,ll}\Phi_{,mm} \,.
\end{align}
We note that the time transformation $\mathcal{L}$ in (\ref{xfn2}) and the use of $\tau$ instead of the proper time $t$ is crucial for obtaining the correct Newtonian description at second order \citep{Rampf:2012pu}. This is an element missing from earlier relativistic ``Lagrangian'' approaches such as \citep{Matarrese:1995sb} or \citep{Stewart:1994wq}. Our choice of the initial values of $J(a)$ and $L(a)$ mentioned above and the initial (almost) uniform density time-slice mean that the particles are initially distributed in a uniform manner on the grid with zero peculiar velocities and the two frames coincide.\footnote{For an alternative view see \citep{Rampf:2013ewa}.} As time progresses, the particles are displaced from their initial positions with the time functions in their growing mode. In a sense we have smoothly matched a long wavelength comoving frame to the Newtonian one at the beginning of the simulation.

How does the above expression compare to the Newtonian results produced by N-body codes? The first term on the rhs of (\ref{x3}) is precisely the ZA. Note that this is the full result coming from GR at this order in the potential and thus the Zel'dovich approximation already captures the leading order (scalar) GR effects. Since the ZA is a Newtonian result and also a very accurate description practically until shell crossing, it is clear that Newtonian dynamics provide the correct particle trajectories even on very large scales. The last term in the first line is exactly the next order result in Newtonian 2LPT \citep{Buchert:1993xz, Stewart:1994wq}. Thus the gradient expansion, properly transformed, reproduces the Newtonian expressions on sub-horizon scales, including the non-local terms. The next two terms are absent in Newtonian theory and represent relativistic corrections at next-to-leading order. Further inspection reveals that the terms of higher order in $\Phi$ scale differently with wavenumber. Counting spatial gradients, the second order Newtonian terms (second r.h.s term in the first line of (\ref{x3})) are enhanced by two spatial derivatives and dominate over the relativistic terms at small scales. This is of course not surprising since Newtonian behaviour is expected to dominate on short scales. The relativistic terms on the other hand are of zeroth order in spatial derivatives and begin to become important on scales of the order of the Hubble radius, eventually dominating over the second order Newtonian terms on large scales. In the next section we quantify these qualitative remarks.

\section{Simulations of large scale motions}

In order to assess the importance of the relativistic terms obtained from GEM (\ref{metric-phi}) we create density fields by displacing the particles according to (\ref{x2}) and (\ref{x3}). To present the effects of the different terms we label them as follows: \{1\} is the Zel'dovich approximation (ZA) alone, \{2\} also includes the second order Newtonian term (2LPT) and \{3\} and \{4\} represent the two relativistic terms of the second and third lines in (\ref{x3}). We then compute three different real-space density fields at redshifts $z=49$ and z=0 using the same realization of the initial conditions $\Phi(\vec k)$ but resulting from displacement of the particles using three different combinations of terms: all of the terms \{1234\}, only ZA and 2LPT \{12\}, and ZA alone \{1\}. At z=49, a typical redshift for starting N-body simulations, the particle horizon is approximately 2.24 Gpc in comoving scale, while at z=0 it is approximately 14.4 Gpc (taking $H_0$= 68 km sec$^{-1}$ Mpc$^{-1}$ \citep{Ade:2013zuv}).

We are interested in two orthogonal qualities of the realizations: firstly the overall variance of the density contrast on different scales and secondly the spatial correlations of structures, {\em i.e.} whether lumps or voids are located at the same place. The former is quantified by the usual power spectrum $P(k)$ which is however insensitive to the spatial positions of features in the realization. The latter is quantified by a cross-correlation~\citep{Coles:1992vr}
\begin{align}
C=\frac{\left<XY\right>}{\sqrt{\left<X^2\right>\left<Y^2\right>}},\label{eq:coleslaw}
\end{align}
between two realizations $X$ and $Y$, where $X$ and $Y$ denote distinct versions of $\delta \rho(\vec x)$ computed using the different terms in (\ref{x3}) and brackets denote averages over the whole volume of the simulations. The quantity (\ref{eq:coleslaw}) is equal to unity in case of full agreement of spatial positions of objects in the $X$ and $Y$ fields and equal to zero when there is no correlation. Note that (\ref{eq:coleslaw}) is insensitive to overall differences in amplitude or the variance.

We use grids of $N=120^3$ points corresponding to 20 different values for the distance $L_{\rm cell}$ between the grid points. These 20 values of $L_{\rm cell}$ are logarithmically spaced between 1 and 1000 Mpc. For each value of physical distance for $L_{\rm cell}$, we create 100 independent realizations of the initial gaussian random potential $\Phi(\vec x)$ with a power spectrum generated using CAMB \citep{Lewis:1999bs}. We then average over the results of these different realizations and obtain error bars by computing the sample variance.

\begin{figure*}
\includegraphics[width=0.45\textwidth]{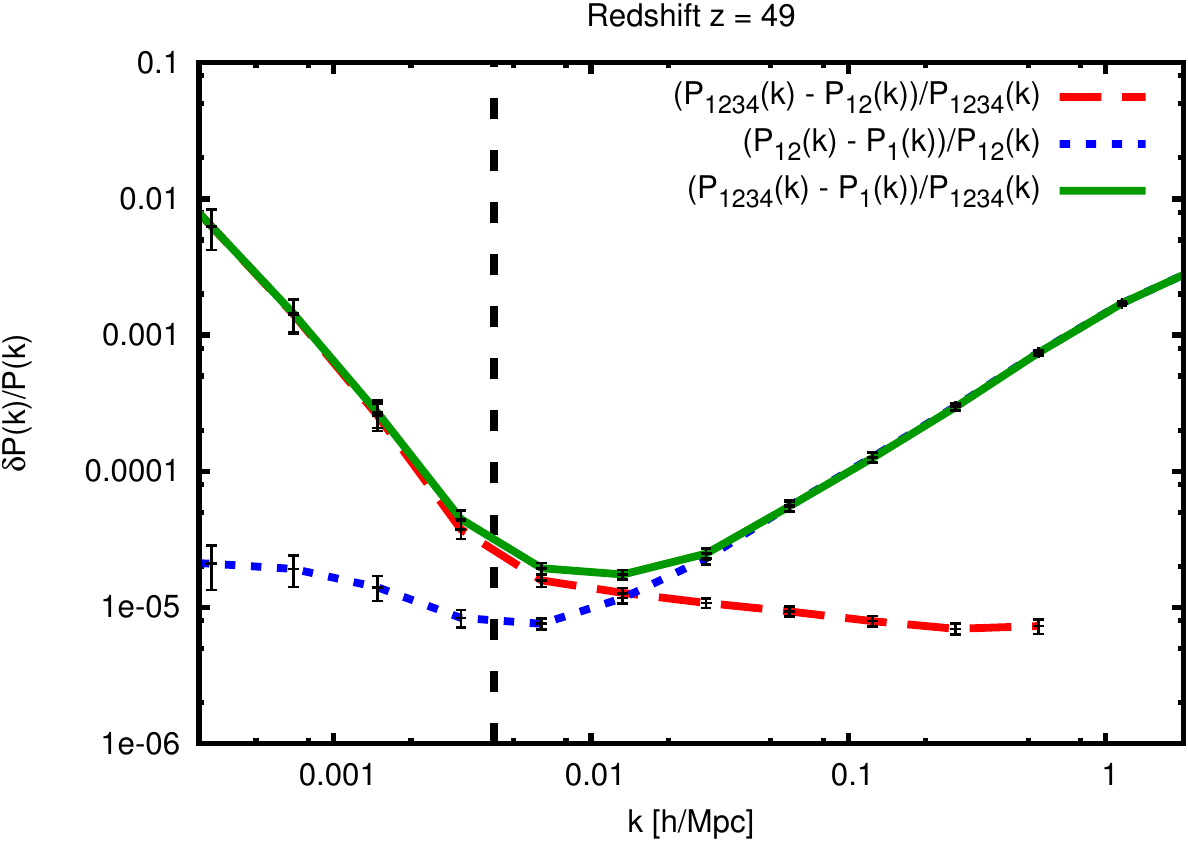}\hspace{0.07\textwidth}
\includegraphics[width=0.45\textwidth]{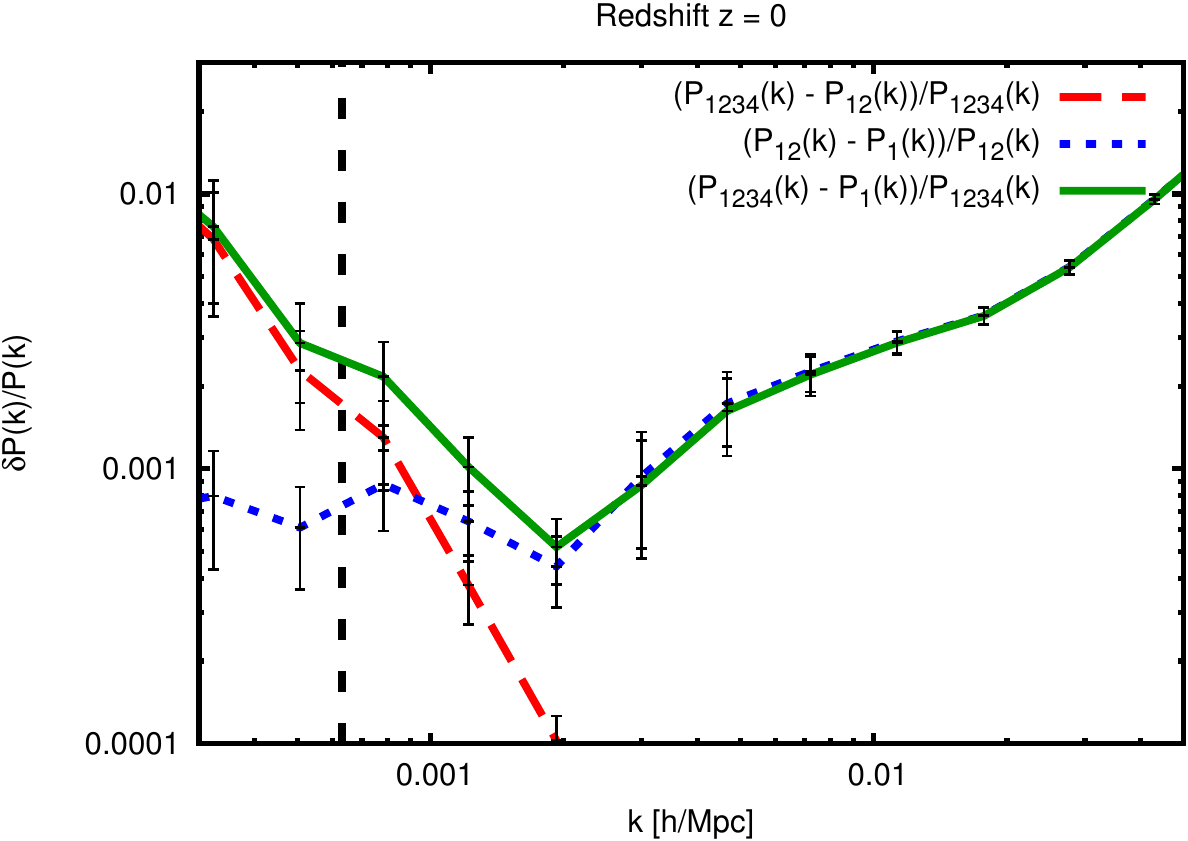}
\caption{The normalized difference in power spectra P(k) for different contributions to the particle displacement ({\em left:} at $z=49$, {\em right:} at $z=0$). The solid (green) line shows the deviation of the Zel'dovich Approximation (ZA, labeled $\{1\}$) from the full Gradient Expansion Metric (GEM, labeled $\{1234\}$), the long dashed line (red) shows the deviation of 2nd-order Lagrangian Perturbation Theory (2LPT, labeled $\{12\}$) from GEM, while the short dashed line (blue) shows the deviation of ZA from 2LPT.
We see that on small scales the correction to ZA is entirely due to 2LPT contributions, while on scales larger than the particle horizon (indicated by a vertical dashed line) it is entirely due to the relativistic terms in GEM, showing the 2LPT terms to be subdominant on such scales. The values are obtained by combining the power spectra of all simulations, with 20 different values for $L_{\rm cell}$, logarithmically spaced between 1 and 1000 Mpc, and 100 realisations per choice of $L_{\rm cell}$, hence a total of 2000 realizations. The error bars are obtained by computing the sample variance at a particular wavenumber for each realization of given size $L_{\rm box}$, and then by propagating these errors to the binned results of the combined realizations.  Note that the range of the vertical scale is different between the left and the right pane, such that error bars are in fact relatively not much different between them. redshifts.}\label{fig:pofk}
\end{figure*}

In Figure~\ref{fig:pofk} we show the relative difference in the power between GEM and ZA ($\{1234\}-\{1\}$), GEM and 2LPT ($\{1234\}-\{12\}$) and 2LPT and ZA ($\{12\}-\{1\}$) at z=49 and z=0. We compute the power spectrum $P(k)$ from all realizations and bin the resulting spectra into one function. The difference between ZA and GEM on small scales is entirely attributed to the difference between ZA and 2LPT, showing that on these scales 2LPT suffices and the relativistic terms are negligible. On large scales the difference between the full result and the ZA is dominated by the relativistic terms in GEM. The crossover between the two regimes happens at scales slightly smaller than the particle horizon scale at these epochs (at z=49, $k_{\rm horizon}\simeq 2.8 \times 10^{-3}$ Mpc$^{-1}$, and at z=0, $k_{\rm horizon}\simeq 0.44\times 10^{-3}$ Mpc$^{-1}$), indicated by a vertical dashed line. This agrees with a qualitative comparison of the different terms in (\ref{x3}).

\begin{figure*}
\includegraphics[width=0.45\textwidth]{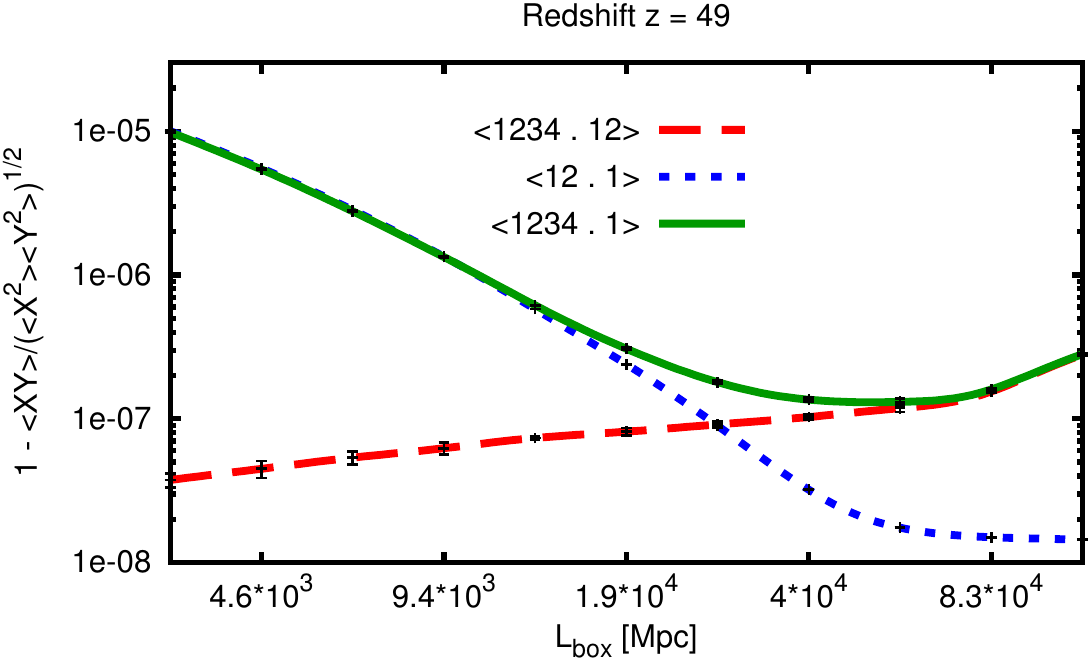}\hspace{0.07\textwidth}
\includegraphics[width=0.45\textwidth]{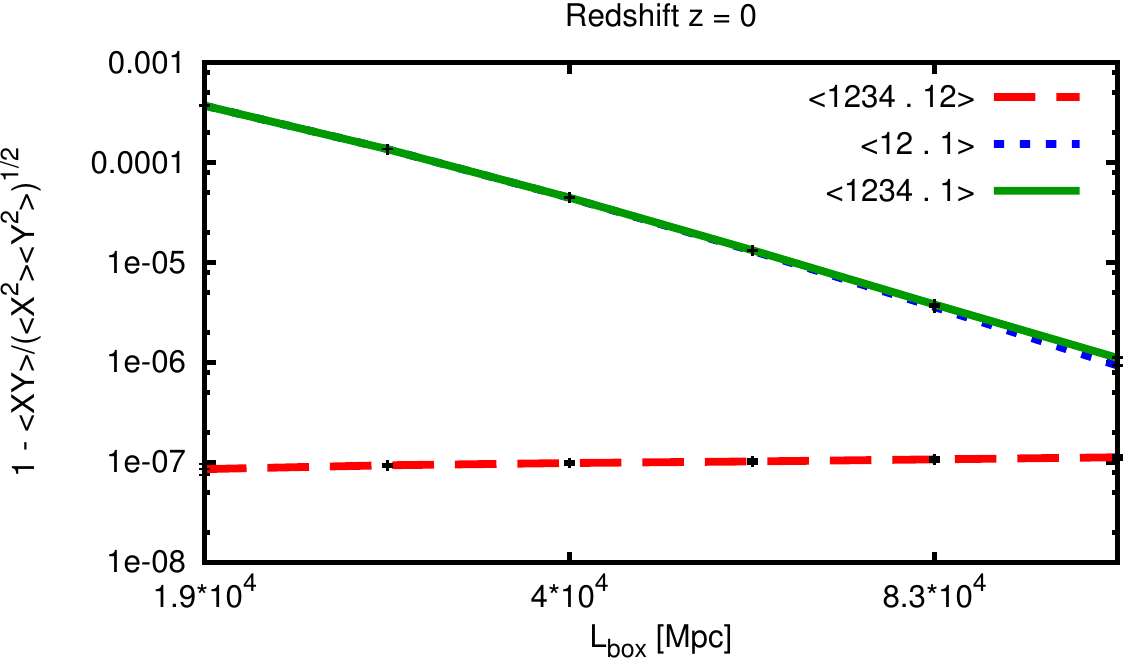}
\caption{The deviation in spatial correlation between two distributions, $1-\frac{\left<XY\right>}{\sqrt{\left<X^2\right>\left<Y^2\right>}}$. This quantity (insensitive to differences in amplitude or variance) is equal to zero for perfect agreement of the distribution of matter and unity for no correlation. The solid (green) line shows the deviation of the Zel'dovich Approximation (ZA, labeled $\{1\}$) from the full Gradient Expanded Metric (GEM, labeled $\{1234\}$), the thick dashed line (red) shows the deviation of 2nd-order Lagrangian Perturbation Theory (2LPT, labeled $\{12\}$) from GEM, while the thin dashed line (blue) shows the deviation of ZA from 2LPT. On small scales the difference between ZA and GEM is as large as the difference between ZA and 2LPT. On these scales the relativistic terms in GEM are irrelevant. On large scales on the other hand on the other hand the difference between ZA and 2LPT is smaller than the total deviation of ZA from GEM, showing that here the relativistic terms dominate the 2LPT contributions. The set of simulations and the estimation of error bars are as before.}
\label{fig:spacecorr}
\end{figure*}

In Figure~\ref{fig:spacecorr} we plot the correlation $1-C$ from Eq.~\eqref{eq:coleslaw} between GEM and ZA ($\langle1234\cdot\,1\rangle$), GEM and 2LPT ($\langle1234\cdot\,12\rangle$) and 2LPT and ZA ($\langle12\cdot\,1\rangle$) at $z=49$ and $z=0$. A similar behavior is observed. For simulations in small boxes, the difference between ZA and GEM is as large as the difference between ZA and 2LPT, meaning that on these scales the relativistic terms in GEM are irrelevant for the correlations (\ref{eq:coleslaw}). For large simulations on the other hand, the difference between ZA and 2LPT is smaller than the total deviation of ZA from GEM, showing that here the relativistic terms dominate over the 2LPT contributions. Here however, the crossover occurs for box sizes significantly larger than the Hubble radius. We thus see that the positions of features in the density field at second order in the gravitational potential are mostly determined by Newtonian terms even for boxes commensurate with the hubble radius. Relativistic terms start becoming important for the particles' spatial positions only at much larger scales.

As a third indicator of the relation between Newtonian and relativistic terms of the particle displacement we show in Figure~\ref{fig:velo2} the average deviations in magnitude of their velocities. Just as the real-space correlations, these values are averages over an entire simulation's box, and the variance in these values is obtained from the 100 realizations performed for each value of $L_{\rm cell}$. Again, for large boxes the velocity corrections are dominated by the GEM terms as opposed to those coming from the 2LPT terms, with the opposite behavior occurring on small scales.

The fact that the quantitative effect on velocities is very similar to the effect on positions, {\em i.e.} GR effects disappearing once a length scale is sufficiently deep inside the horizon, suggests that the relativistic contributions to the velocities `correct' for the relativistic positions. It seems that ignoring all GR contributions and taking 2LPT at both early and late times, leads to the correct late time particle positions and velocities. That is, the GR contributions to the velocities are such that the GR contributions to positions fade out over time \citep{Adamek:2013wja}. We hope to investigate this further in future work.

\begin{figure*}
\includegraphics[width=0.45\textwidth]{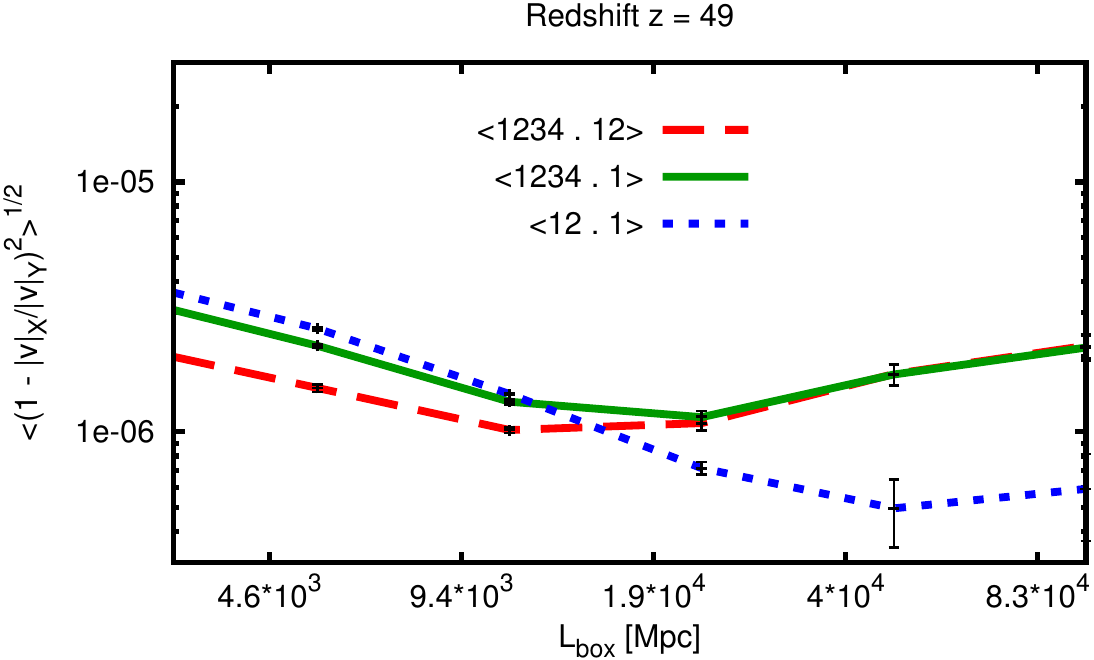}
\hspace{0.07\textwidth}
\includegraphics[width=0.45\textwidth]{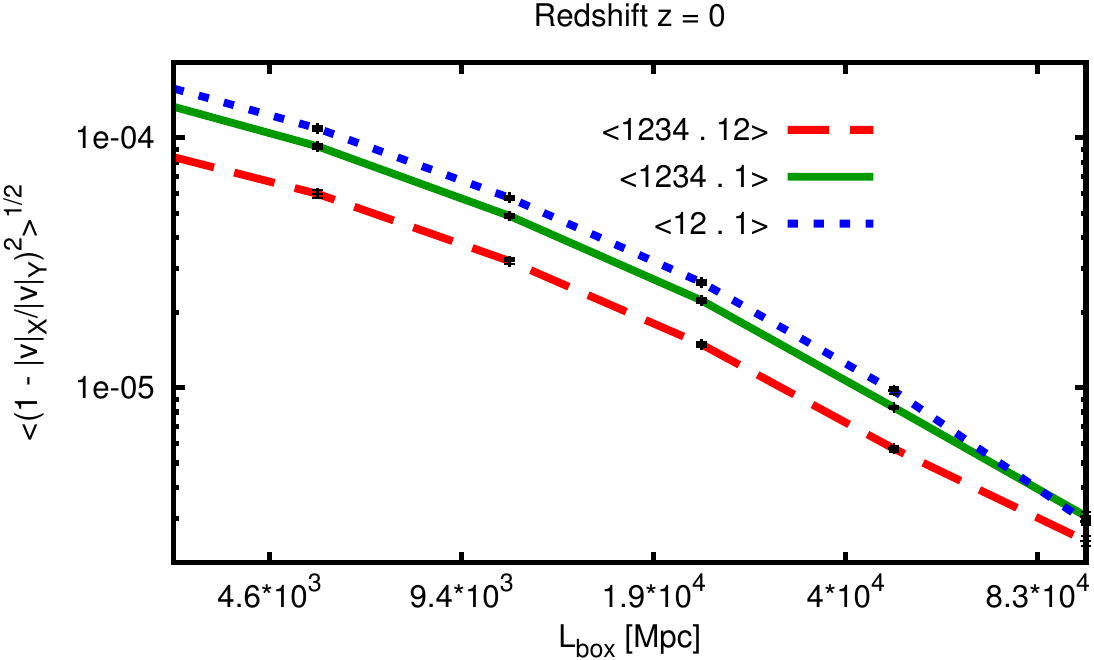}
\caption{Average relative difference in velocity magnitudes. The quantity $(1-\left|\vec v_X\right| / \left| \vec v_Y \right|)$ compares the magnitude of the velocity of one particle at some lagrangian position $\vec q$, between two simulations with the same random seed but including different combinations of terms $\{1234\}$.  The quantity $\left<(1-\left|\vec v_X\right| / \left| \vec v_Y \right|)\right>$ is averaged of all lagrangian coordinates in one realization. The measured values in this figure each come from averaging over 100 realizations with a given $L_{\rm cell}$.  As with the power spectra and the real space correlations, the velocities become affected by the general relativistic corrections for large simulations, while the Newtonian terms dominate the second order displacements for smaller boxes.\label{fig:velo2}}
\end{figure*}

\section{Conclusion and Discussion}

Motivated by the increasing size of cosmological N-body simulations, which can now encompass the whole of the observable universe, we have investigated the long wavelength effects of General Relativity on the motion of CDM particles in $\Lambda$CDM cosmology. We took the view that the coordinates employed in N-body simulations, the Euclidean grid on which particles move and the clock that defines successive moments in the simulation, are to be identified with the coordinates of a Newtonian system in which the metric takes the form (\ref{Newton-metric}) - they correspond to the choice of the Newtonian gauge. On the other hand, the long wavlength behaviour of CDM under the action of gravity is conveniently described in a synchronous comoving frame with the metric (\ref{metric-phi}). The transformation between the two provides the trajectory of the particles in the Newtonian frame, including possible relativistic corrections. This procedure is akin to a Lagrangian-to-Eulerian transformation in a relativistic setting and thus appropriate for understanding particle trajectories in the Newtonian frame even on scales approaching or exceeding the horizon. These are the scales where one might question the validity of purely Newtonian simulations.

The leading order GR solution exactly coincides with the Zel'dovich approximation \citep{Rampf:2012pu, Russ:1995eu}. Since the latter is also the leading order solution to the Newtonian equations, we see that using newtonian simulations provides the correct particle motion to leading order, even for box sizes exceeding the horizon. Differences do show up at next-to-leading order. In this paper we have quantified these differences and the scale at which the crossover occurs between second order Newtonian and second order relativistic terms at two different redshifts - see figures \ref{fig:pofk}, \ref{fig:spacecorr} and \label{fig:velo2}. We have thus explicitly shown for the first time that for scales up to the horizon 2LPT suffices as a correction to the Zel'dovich approximation. On super horizon scales, scalar relativistic corrections dominate over the Newtonian 2LPT terms. This behaviour is evident in the amplitude of perturbations, the spatial correlations of structures as well as the magnitudes of the velocities. We found that the crossover scale is larger for the last two quantities compared with the power spectrum for which the crossover occurs approximately at horizon scales.

Upcoming simulations could conceivably go up to comoving scales a few times larger than the current particle Horizon. From figures \ref{fig:pofk} and \ref{fig:spacecorr} we see that at z=49 the dominant relativistic corrections to the power spectrum on such scales are at most of the order of $10^{-3}$, while the correlations of spatial features are affected by the dominant relativistic terms at a level below $10^{-6}$. Similar conclusions hold for super-horizon scales at z=0. Hence, setting up initial conditions using the Zel'dovich approximation or, more generally, using Newtonian physics up to very large scales gives essentially the correct particle trajectories to the accuracy stated.


We stress that these statements refer only to the positions of particles with respect to the coordinates of a Newtonian reference frame. Inferring other quantities such as the density from the simulation output  requires more care. One might be tempted to insert the transformation (\ref{xfn1}) and (\ref{xfn2}) into (\ref{density-pert}) and expand to the desired order. This simply returns the density perturbation in the Newtonian gauge:
\beq\label{density-pert-newt}
\frac{\delta\rho}{\bar{\rho}}\left(\tau,\vc{x}\right)\simeq -\frac{\lambda(\tau)}{a^2(\tau)}\frac{10}{3}\nabla^2_{\vc{x}}\Phi(\vc{x}) + 10 H(\tau)J(\tau)\Phi(\vc{x})\,,
\eeq
which is also the expression given in the ``dictionary'' of \citep{Green:2011wc}. Note that this is a non-gauge-invariant quantity. Although not being gauge invariant is not necessarily a condemning attribute, it can lead to interpretational difficulties on large scales. One drawback of (\ref{density-pert-newt}) is that it contains a term proportional to $\Phi$ that is not suppressed by spatial gradients and thus leads to divergent contributions to the power-spectrum on large scales - see eg figure 1 of \citep{Flender:2012nq}.\footnote{An extra term proportional to $\Phi$, resulting form an initial displacement of coordinates to match the Newtonian gauge expression for the density was also suggested in \citep{Chisari:2011iq}, see also \citep{Rampf:2012pu}. In fact such a term can be absorbed on the initial hypersurface by a remaining gauge freedom in the comoving synchronous gauge, see the appendix of \citep{Russ:1995eu}.} To obtain a gauge invariant quantity at leading order one can subtract from (\ref{density-pert-newt}) the term $-3Hv$, where $v$ is the velocity potential \citep{Bardeen:1980kt}. This eliminates the term proportional to $\Phi$ as can be easily seen from eqs (\ref{coeffs}), (\ref{x2}) and (\ref{x3}).

In this paper, we have automatically used a gauge invariant quantity for the density to all orders by considering the synchronous frame quantity (\ref{density}), with the identification of the numerical values of $t$ and $\vc{q}$ with $\tau$ and $\vc{x}$, ie \emph{without} using (\ref{xfn1}) and (\ref{xfn2}). At leading order this gives (\ref{density-pert}). This amounts to identifying the synchronous frame density, a physically well-defined quantity, as the relevant gauge invariant observable in the Newtonian frame. We have set $\frac{\delta\rho_i}{\rho_i}=0$ which recovers the growing mode.

Let us end by noting that the  considerations of this work are necessary for properly interpreting the output of an N-body simulation. However, in order to compute what an observer will actually infer from the perturbations on such scales by receiving light, one should further rely on a GR framework that treats the propagation of light rays through the resulting inhomogeneous spacetime - see for example \citep{Yoo:2009au,Yoo:2010ni, LopezHonorez:2011cy, Bonvin:2011bg, Bruni:2011ta}.

\section{Acknowledgements}
We would like to thank Filippo Vernizzi and Pier-Stefano Corasaniti for useful discussions and Matthias Bartelmann for his comments on the manuscript. GR  is supported by the Gottfried Wilhelm Leibniz programme of the Deutsche Forschungsgemeinschaft (DFG). WV is supported by a Veni research grant from the Netherlands Organization for Scientific Research (NWO).

\bibliographystyle{mn2e}
\bibliography{grcorr_biblio}

\end{document}